\documentclass[
prl,               
twocolumn,         
superscriptaddress,%
longbibliography,  
floatfix,          
]{revtex4-2}

\usepackage{graphicx}
\usepackage{amsmath}
\usepackage{amssymb}
\usepackage{bm}
\usepackage{siunitx}
\usepackage{xcolor}
\usepackage{mhchem}
\usepackage{url}
\usepackage{hyperref}
\usepackage[utf8]{inputenc}
\usepackage{ragged2e}
\usepackage[capitalize,nameinlink]{cleveref}
\usepackage[labelfont=bf,textfont=normal]{caption}
\usepackage{float}
\floatstyle{plaintop}
\restylefloat{table}

\hypersetup{
    colorlinks=true,
    citecolor=blue,
    linkcolor=blue,
    urlcolor=blue,
}

\begin{document}

\title{Intrinsic anomalous Hall response in the bilayer kagome ferromagnet Co$_3$Sn}

\author{Yuqi Qin}
\thanks{These authors contributed equally.}
\affiliation{Department of Physics, The Hong Kong University of Science and Technology, Clear Water Bay, Kowloon, Hong Kong SAR}
\author{Soumya Sankar}
\thanks{These authors contributed equally.}
\affiliation{Department of Physics, The Hong Kong University of Science and Technology, Clear Water Bay, Kowloon, Hong Kong SAR}
\author{Xingkai Cheng}
\affiliation{Department of Physics, The Hong Kong University of Science and Technology, Clear Water Bay, Kowloon, Hong Kong SAR}
\author{Yifan Jiang}
\affiliation{Department of Physics, The Hong Kong University of Science and Technology, Clear Water Bay, Kowloon, Hong Kong SAR}
\author{Shiming Lei}
\affiliation{Department of Physics, The Hong Kong University of Science and Technology, Clear Water Bay, Kowloon, Hong Kong SAR}
\author{Junwei Liu}
\affiliation{Department of Physics, The Hong Kong University of Science and Technology, Clear Water Bay, Kowloon, Hong Kong SAR}
\author{Berthold J\"{a}ck}
\email{bjaeck@ust.hk}
\affiliation{Department of Physics, The Hong Kong University of Science and Technology, Clear Water Bay, Kowloon, Hong Kong SAR}
\email{bjaeck@ust.hk}

\date{\today}

\begin{abstract}
Transition-metal kagome magnets provide a rich platform for investigating the interplay between layer stacking, magnetic order, and band topology. Here, we report the molecular beam epitaxy and experimental investigation of high-quality thin films of the kagome metal Co$_3$Sn, which has not been synthesized in bulk form yet. Structural and chemical analyses confirm a hexagonal lattice structure ($P6_3/mmc$) composed of direct A-B stacked Co$_3$Sn kagome bilayers. Magnetometry reveals robust easy-plane ferromagnetism with a Curie temperature exceeding $300\,\text{K}$. Magneto-transport measurements demonstrate metallic behavior (carrier density $n\approx5.01\times10^{22}\,\text{cm}^{-3}$) alongside a temperature-independent anomalous Hall conductivity of $\sigma_{\rm AHE}\approx90\,\Omega^{-1}\cdot{\rm cm}^{-1}$, extending from $2\,\text{K}$ up to room temperature. Results from first-principles density functional theory calculations attribute this anomalous Hall response to intrinsic Berry curvature hotspots near the Fermi level in the spin-split band structure. Our results establish Co$_3$Sn as a room-temperature kagome ferromagnet and highlight the impact of the layer stacking sequence on the material properties of kagome metals from the CoSn-family.
\end{abstract}

\maketitle

\section{INTRODUCTION}
Transition-metal-based magnetic kagome compounds have captured widespread attention in recent years owing to their unconventional topological responses. The kagome lattice, which consists of a two-dimensional network of corner-sharing triangles, has long been recognized as a fertile ground for exotic quantum behavior, including geometric frustration, flat bands, Dirac cones, and Weyl fermions, and it can host cascades of phase transitions due to correlations in topological flat bands~\cite{yin2018giant,chen2023visualizing,kang2020topological,ghimire2020topology,chen2026nematic,nakatsuji2015large}. When strong spin–orbit coupling and magnetism are present, these materials exhibit notable transport responses such as a large anomalous Hall effect (AHE)~\cite{liu2018giant,wang2018large}, the nonlinear anomalous Hall effect~\cite{sankar2024experimental}, as well as the anomalous in-plane Hall effect~\cite{sankar2025room, sankar2026distinguishing} that are intimately linked to the Berry curvature of the electronic band structure~\cite{nagaosa2010anomalous,xiao2010berry,zhang2023higher}. Hence, these anomalous transport responses make kagome magnets suitable platforms for potential applications in the areas of spin orbit logic~\cite{manipatruni2019scalable}, energy harvesting~\cite{onishi2024high}, or magnetic field sensing~\cite{ni2016ultrahigh}. 

\begin{figure}[htbp]
\centering
\includegraphics[width=\columnwidth]{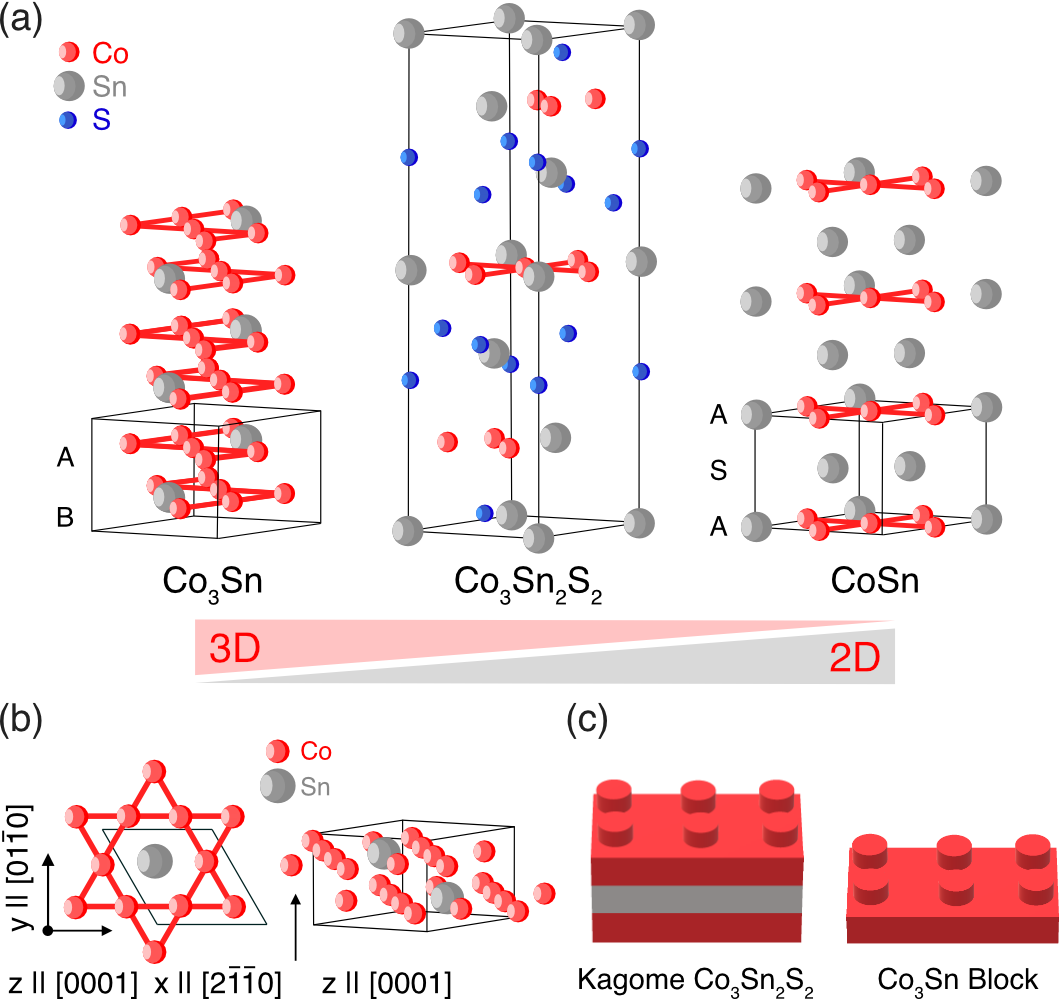}
\caption{\justifying\textbf{Co$_3$Sn: A kagome Lego}. (a) Shown is the alternating arrangement of kagome Co$_3$Sn and stanene (Sn$_2$) layers, which facilitates the transition from a 3D compound to a quasi-2D compound CoSn. (b) Shown is the crystal structure of a Co$_3$Sn layer (left) and the 3D crystal structure of Co$_3$Sn. (c) Shown is the formation of Co$_3$Sn$_2$S$_2$ kagome lattices from the Co$_3$Sn kagome block.}
\label{fig1}
\end{figure}

The prototypical ferromagnetic Weyl semimetal Co$_3$Sn$_2$S$_2$ has been established as a cornerstone of topological magnetism over recent years, predominantly owing to its unprecedentedly large intrinsic AHE Hall conductivity exceeding $10^3\,\Omega^{-1}\cdot {\rm cm}^{-1}$~\cite{liu2018giant,wang2018large}. This compound belongs to the larger family of transition-metal kagome metals [Fig.~\ref{fig1}(a)], denoted as \(X_{p}Y_{q}\) (where \(X = \text{Fe, Co, Ni}\) and \(Y = \text{Sn, In, Ge}\)) that has recently emerged as a fertile ground for exploring the intricate interplay between topology, electron correlations, and magnetic order~\cite{liu2018giant, wang2018large, ye2018massive, kang2020dirac, kang2020topological, sankar2024experimental, chen2026nematic}. A shared building block of this material family is a quasi-2D Co$_3$Sn kagome layer [Fig.~\ref{fig1}(b)], whose out-of-plane layer stacking was found to profoundly influence the electronic and magnetic material properties. In paramagnetic CoSn, the separation of the Co$_3$Sn layers by stanene (Sn$_2$) layers preserves the quasi-2D nature of the electronic band structure~\cite{kang2020topological}, resulting topological flat bands that can host strongly-interacting states at partial flat band filling~\cite{chen2023visualizing, chen2026nematic}. A more complex out-of-plane coordination of the Co$_3$Sn planes in Co$_3$Sn$_2$S$_2$ leads to the formation of out-of-plane ferromagnetism as well as topological Weyl points in the electronic band structure~\cite{liu2018giant,wang2018large}. 

Another stacking possibility, which is the A-B bilayer stacking of the Co$_3$Sn layers [Fig.~\ref{fig1}(b)], has remained unexplored so far, owing to the difficulty in synthesizing metastable Co$_3$Sn. Previous studies of the FeSn series have shown that this A-B stacking induces a transition from massive Dirac fermions to topological Weyl points in the electronic band structure of Fe$_3$Sn$_2$~\cite{ye2018massive} and Fe$_3$Sn~\cite{belbase2023large, prodan2023large, sankar2025room}, respectively. Hence, a natural question arises how the transition from spaced-out kagome layers in Co$_3$Sn$_2$S$_2$ to the direct A-B layer stacking in Co$_3$Sn influences the electronic and magnetic properties, potentially leading to unique experimental phenomenologies. Here, we present the first experimental realization of Co$_3$Sn using thin-film growth with molecular beam epitaxy (MBE). Combining detailed materials characterization with first-principle calculations, we find that Co$_3$Sn exhibits in-plane ferromagnetism and a sizeable anomalous Hall responses that arises from Berry curvature in the spin-split band structure.

\section{METHODS}
\label{sec:methods}

\subsection{Molecular beam epitaxy growth}

Epitaxial thin films of Co$_3$Sn were grown on (0001)-oriented Al$_2$O$_3$ substrates using a home-built MBE system with a base pressure below $5\times10^{-10}$ Torr. Prior to Co$_3$Sn growth, a Pt(111) layer of 5 nm thickness was grown~\cite{cheng2022atomic,sankar2025room}. High-purity Co (99.99\%) and Sn (99.999\%) were evaporated from Knudsen effusion cells. The substrate temperature was maintained at 200–220\,$^\circ$C, as determined by a pyrometer. The Co and Sn fluxes were calibrated using a beam flux monitor. The growth rate was approximately 1.6\,Å/min with a Co:Sn flux ratio of 1:1.3. After growth, the resulting films were continuous, with a thickness of about 50\,nm.

\subsection{Structural and chemical characterization}

The crystal structure and phase purity were examined by X-ray diffraction (XRD) using a Panalytical Empyrean MultiCore PIX3D diffractometer (Malvern Panalytical B.V.) with Cu K$\alpha$ radiation ($\lambda = 1.5406$ \AA). Data were collected over $10^\circ \leq 2\theta \leq 100^\circ$ with a step size of $0.02^\circ$. X-ray photoelectron spectroscopy (XPS) measurements were conducted with a Kratos Axis Supra+ spectrometer (Kratos Analytical Ltd.) equipped with a monochromatic Al K$\alpha$ source ($h\nu = 1486.6$\,eV). The binding energy scale was calibrated against the C 1s peak at 284.8\,eV. Both survey scans and high-resolution core-level spectra for Co 2p and Sn 3d were acquired.

\subsection{Magnetometry and and magneto-transport measurements}

The sample magnetization was measured using a Quantum Design MPMS3 SQUID magnetometer over the temperature range 2–400 K and magnetic fields up to 7 T. For magneto-transport measurements, the as-grown films were patterned into a circular Hall bar geometry via photolithography and Ar$^+$ ion milling. The longitudinal ($\rho_{xx}$) and transverse resistivity ($\rho_{xy}$) were measured in a Quantum Design Physical Property Measurement System (PPMS) using a low-frequency AC excitation bias current (19.375\,Hz, 700 $\mu$A) with standard lock-in techniques.

\subsection{First-principles calculations}

First-principles calculations of the electronic structure of Co$_3$Sn were performed using the density functional theory framework as implemented in the Vienna {\em ab initio} simulation package~\cite{kresse1996efficiency, kresse1996efficient}. The projector-augmented wave potential was adopted with the plane-wave energy cutoff set to 600\,eV (convergence criteria $10^{-6}\,$eV). The exchange-correlation functional of the Perdew–Burke–Ernzerhof type was used~\cite{perdew1996generalized} with a 7$\times$7$\times$9 gamma-centered Monkhorst–Pack mesh. We constructed the Hamiltonian of a Wannier tight-binding model using the {\em WANNIER90} interface~\cite{mostofi2008wannier90}, including the Co $d$-and Sn $p$-orbitals. We then calculated the anomalous Hall conductivity from this Wannier model using a 201$\times$201$\times$151 k-mesh.

\section{RESULTS}
\label{sec:results}


\subsection{Characterization of crystal structure with RHEED and X-ray Diffraction}

Co$_3$Sn crystallizes in a hexagonal lattice structure ($P6_3/mmc$, $a=b=5.32\,{\rm Å}$, $c=4.23\,{\rm Å}$) that it shares with its iso-structural siblings Fe$_3$Sn and Ni$_3$In~\cite{belbase2023large, prodan2023large, ye2024hopping}. Each kagome layer in the crystallographic $ab$-plane comprises a kagome net composed of Co ions as well as a coordinating Sn ions at the hexagon center [Fig.~\ref{fig1}(b)]. The primitive unit cell of Co$_3$Sn [Fig.~\ref{fig1}(b)] is composed of two kagome bilayers stacked along the crystallographic [0001]-direction.

The MBE of the Co$_3$Sn films was monitored using in-situ reflection high-energy electron diffraction (RHEED)~\cite{ichimiya2004reflection,cho1975molecular}. The RHEED patterns of the as-grown films [Fig.~\ref{fig2}(a)] exhibit sharp streaky features along both the [1$\bar{1}$00] and [11$\bar{2}$0] azimuths, indicating a smooth film surface resulting from a layer-by-layer growth mode. The observation of these streaky patterns throughout the growth process confirms the high crystalline quality and excellent surface morphology of the Co$_3$Sn films.

The XRD spectrum of a $\theta-2\theta$ scan [Fig.~\ref{fig2}(b)] exhibits several diffraction peaks that can be fully indexed to the crystalline structure of Co$_3$Sn. This observation demonstrates the successful synthesis of single-phase Co$_3$Sn films [large range XRD spectrum is shown in the supplementary materials~\cite{SI}]. The refined lattice parameter $c = 4.24$ \AA  is in excellent agreement with theoretical predictions from material databases~\cite{jain2013commentary}. Furthermore, the rocking curve for the (001) reflection obtained from a $\omega$-scan [inset of Fig.~\ref{fig2}(b)] exhibits a full width at half maximum (FWHM) of $0.007^\circ$, indicating exceptional thin film orientation and low mosaic disorder. The in-plane XRD $\phi$-scan at the $2\theta$--$\chi$ angle of $61.2^\circ$, corresponding to the Co$_3$Sn reflection, reveals that the Co$_3$Sn, Pt, and Al$_2$O$_3$ layers are structurally aligned, indicating a successful thin film epitaxy as well as the absence of structural defects, such as disoriented crystalline domains.

\subsection{Chemical composition and magnetism}
\begin{figure}[htbp]
\centering
\includegraphics[width=\columnwidth]{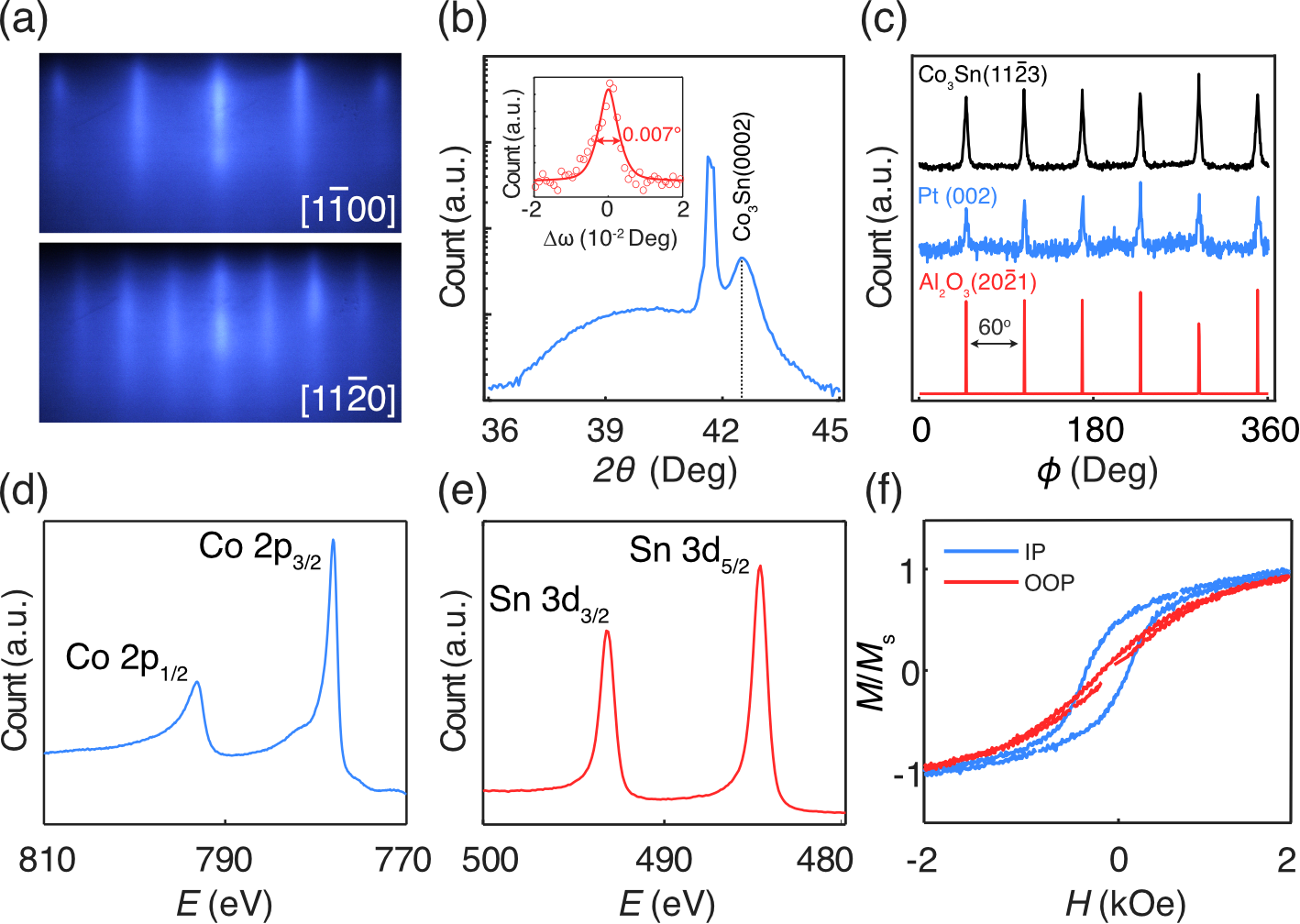}
\caption{\justifying\textbf{Structural and magnetic characterization of the in-plane kagome ferromagnet Co$_3$Sn.} (a) Shown are the RHEED patterns of the Co$_3$Sn thin film after growth along the [1$\bar{1}$00] (top) and [11$\bar{2}$0] (bottom) azimuths. (b) Shown is the X-ray diffraction pattern of a Co$_3$Sn thin film, exhibiting diffraction peaks corresponding to Al$_2$O$_3$ (0006) and Co$_3$Sn (0002) at 41.7$^\circ$ and 42.6$^\circ$, respectively. (c) Shown is the $\phi$-scan spectrum of the Co$_3$Sn thin film. (d) Shown is the XPS core-level spectrum of Co 2p, revealing the characteristic Co 2p$_{3/2}$ and Co 2p$_{1/2}$ peaks along with their shake-up satellites. (e) Shown is the XPS core-level spectrum of Sn 3d, displaying the Sn 3d$_{5/2}$ and Sn 3d$_{3/2}$ peaks. (f) Shown is the magnetization $M$ of the Co$_3$Sn film measured at 300~K for both in-plane (IP) and out-of-plane (OOP) field orientations of the magnetic field $H$. $M$ is normalized to its saturation value $M_s$ at $H=2\,$kOe.}
\label{fig2}
\end{figure}

We further performed a chemical characterization of the as-grown thin films. The XPS core-level spectra, shown in Fig.~\ref{fig2}(d) and (e), reveal Co 2p$_{3/2}$ and Co 2p$_{1/2}$ peaks at 778.2 eV and 793.3 eV, respectively, accompanied by characteristic shake-up satellites typical of metallic Co. The Sn 3d$_{5/2}$ and 3d$_{3/2}$ peaks appear at 485.0 eV and 493.4 eV, confirming the expected metallic state~\cite{biesinger2011resolving}. No signals corresponding to oxidized species (CoO or SnO$_2$) are observed, indicating good chemical stability of the thin film stored under ambient conditions.

The magnetic properties of the Co$_3$Sn thin films were characterized at ambient temperature ($T = 300$ K). The external magnetic field $\mu_0 H$ was applied both parallel (in-plane, IP) and perpendicular (out-of-plane, OOP) to the film plane to probe the orientational dependence of the magnetic response. Figure~\ref{fig2}(f) presents the field-dependent normalized magnetization $M(H)/M_s$, where $M_s$ denotes the saturation magnetization. The acquired $M(H)$ isotherms exhibit well-defined hysteresis loops for both field orientations, revealing the presence of magnetic anisotropy. Notably, the IP loop encloses a substantially larger integrated area than the OOP loop. This observation indicates easy-plane ferromagnetism with the magnetization oriented within the crystallographic $ab$-plane [also see Fig.~\ref{fig1}(b)].

\subsection{Low temperature transport and anomalous Hall effect}
\begin{figure}[htbp]
\centering
\includegraphics[width=\columnwidth]{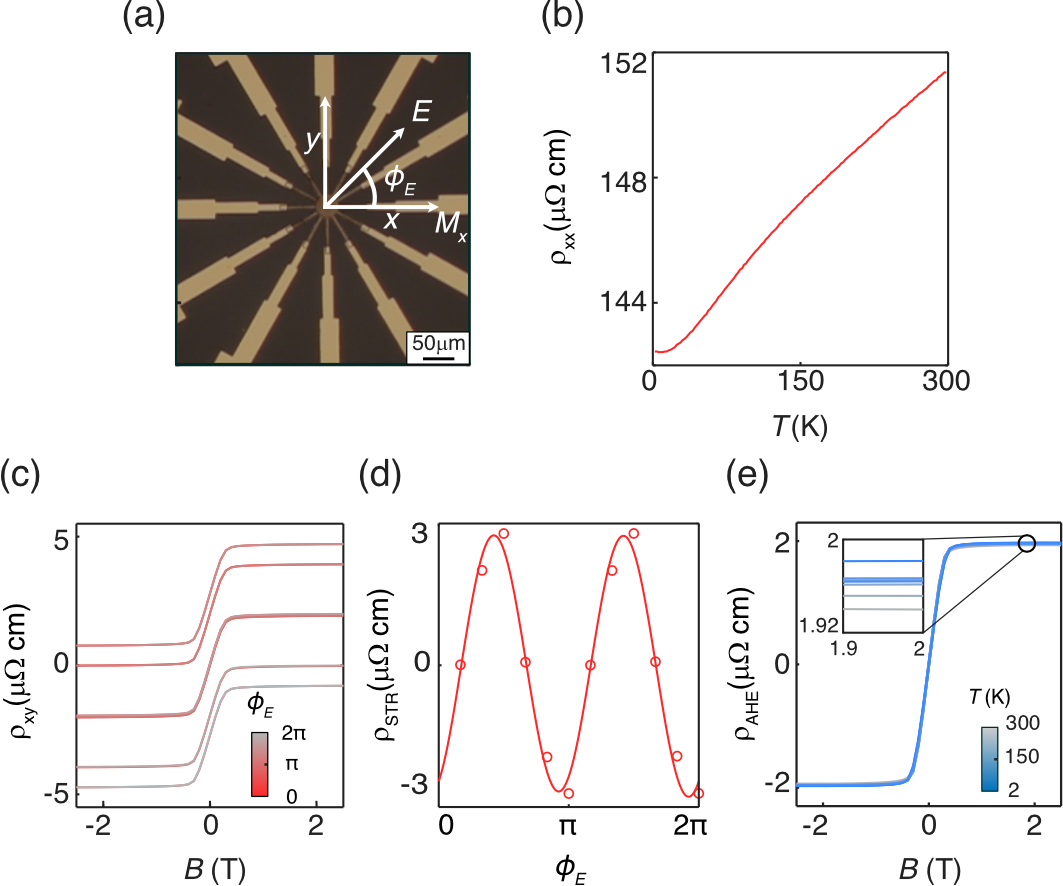}
\caption{\justifying\textbf{Electric transport of in-plane kagome ferromagnetic Co$_3$Sn.} (a) Shown is the optical image of the circular Hall bar device; the magnetization is fixed along $M_x$, and $\phi_E$ is the azimuthal angle between the electric field and the $x$-axis. (b) Shown is the temperature ($T$) dependent longitudinal resistivity ($\rho_{\rm xx}$). (c) Shown is the transverse resistivity ($\rho_{\rm xy}$) as a function of azimuthal angle $\phi_E$. (d) Shown is the dependence of $\rho_{\rm STRS}$ as a function of the azimuthal angle $\phi_E$. (e) Shown is the anomalous Hall resistivity ($\rho_{\rm AHE}$) as a function of the magnetic field $B$ at different indicated temperatures.}
\label{fig3}
\end{figure}

Electric transport measurements were performed on a circular Hall bar device fabricated from the As-grown Co$_3$Sn thin film. The circular device geometry permits angle-dependent measurements as a function of the angle ($\phi_E$) between the bias current and the principal crystal axes, which are suited to determine magnetic anisotropy contributions to $\rho_{xy}$~\cite{sankar2025room,sankar2026distinguishing}. An optical image of the device is shown in Fig.~\ref{fig3}(a). Figure~\ref{fig3}(b) shows the temperature-dependent longitudinal resistivity $\rho_{xx}(T)$. Its room temperature value $\rho_{xx}\approx150\,\mu\Omega$ is comparable to other transition-metal based kagome materials, such as Fe$_3$Sn$_2$~\cite{ye2018massive} and Fe$_3$Sn~\cite{sankar2025room}, and it decreases monotonically from 300~K down to 2~K, indicating a metallic character. The small residual resistivity ratio RRR $= \rho_{300\mathrm{K}}/\rho_{2\mathrm{K}} = 1.08$ suggests that a temperature-independent scattering mechanism dominates the electron transport over electron-phonon scattering ($\propto T$). Given the single-phase nature, high film orientation, and absence of domain defects as determiend with XRD measurements, we suspect that RRR is limited by grain boundary scattering in the thin films~\cite{mar1994grain}.

Next, we consider measurements of the transverse resistivity $\rho_{xy}$ as a function of the magnetic field and $\phi_E$ [Fig.~\ref{fig3}(c)]. We find that $\rho_{xy}$ quickly increases (decreases) with increasing (decreasing) magnetic field before saturating at $B\approx\pm400mT$ where the saturation behavior is consistent with our results from magnetization measurements~[Fig.~\ref{fig2}(f)]. Moreover, we find that $\rho_{xy}$ measured along different in-plane directions exhibits a vertical offset that depends on $\phi_E$ but is independent of $B$. This indicates the presence of anisotropic magneto-resistance (AMR) effects. Therefore, to separate different contributions, $\rho_{xy}$ was decomposed into magnetic field symmetric
\[
\rho_{STR} = \frac{\rho_{xy}(+B) + \rho_{xy}(-B)}{2},
\]
and magnetic-field antisymmetric
\[
\rho_{AS}^{\mathrm{xy}}(B) = \frac{\rho_{xy}(+B) - \rho_{xy}(-B)}{2},
\]
parts, where $\rho_{xy}(+B)$ and $\rho_{xy}(-B)$ are the transverse resistivities measured at positive and negative magnetic fields, respectively. 

The magnetic-field symmetric offset $\rho_{STR}(\phi_E)$ in Fig.~\ref{fig3}(d) reveals a periodic dependence on $\phi_E$ with $\pi$-periodicity. This characteristic dependence of $\rho_{STR}(\phi_E)$ on the direction of the bias current is consistent with the presence of AMR~\cite{mcguire2003anisotropic}. Because AMR requires a low symmetry (not higher than two-fold ($C_2$) out-of-plane rotation), this observation is consistent with the presence of easy-plane ferromagnetism detected in magnetization measurements, which breaks the $C_{3z}$ symmetry within the $ab$-plane.

Next, we consider the magnetic-field antisymmetric part $\rho_{AS}^{\mathrm{xy}}(B)$. To isolate the AHE contribution from the ordinary Hall effect (OHE), which is proportional to $B$, we fit the high-field ($B>1\,$T) linear region of $\rho_{AS}^{\mathrm{xy}}(B)$ to $\rho_{AS}^{\mathrm{xy}}(B)=R_0B+\rho_{\mathrm{AHE}}$, where $R_0$ is the OHE coefficient. The $R_0B$ term was then subtracted from $\rho_{AS}^{\mathrm{xy}}$, yielding the anomalous Hall resistivity $\rho_{\mathrm{AHE}}$. The analysis of the Hall coefficient [see Ref.~\cite{SI}] reveals $p$-type dominated carrier transport at room temperature with a carrier density $n=5.01\times10^{22}\,{\rm cm}^{-3}$. In Fig.~\ref{fig3}(d), we present $\rho_{\mathrm{AHE}}$ measured as a function of temperature $T$ between 300 and 2\,K. Our measurement results show that the saturation value of $\rho_{\mathrm{AHE}}\approx2\mu\Omega\text{cm}$ is nearly temperature-independent, corresponding to an anomalous Hall conductivity $\sigma_{\rm AHE}\approx\rho_{xy}/\rho_{xx}^2\approx90\,\Omega^{-1}{\rm cm}^{-1}$ Measurements on another device that reproduce this observation are presented in the supplementary materials \cite{SI}.

\section{DISCUSSION}
\label{sec:discussion}
\begin{figure}[htbp]
\centering
\includegraphics[width=\columnwidth]{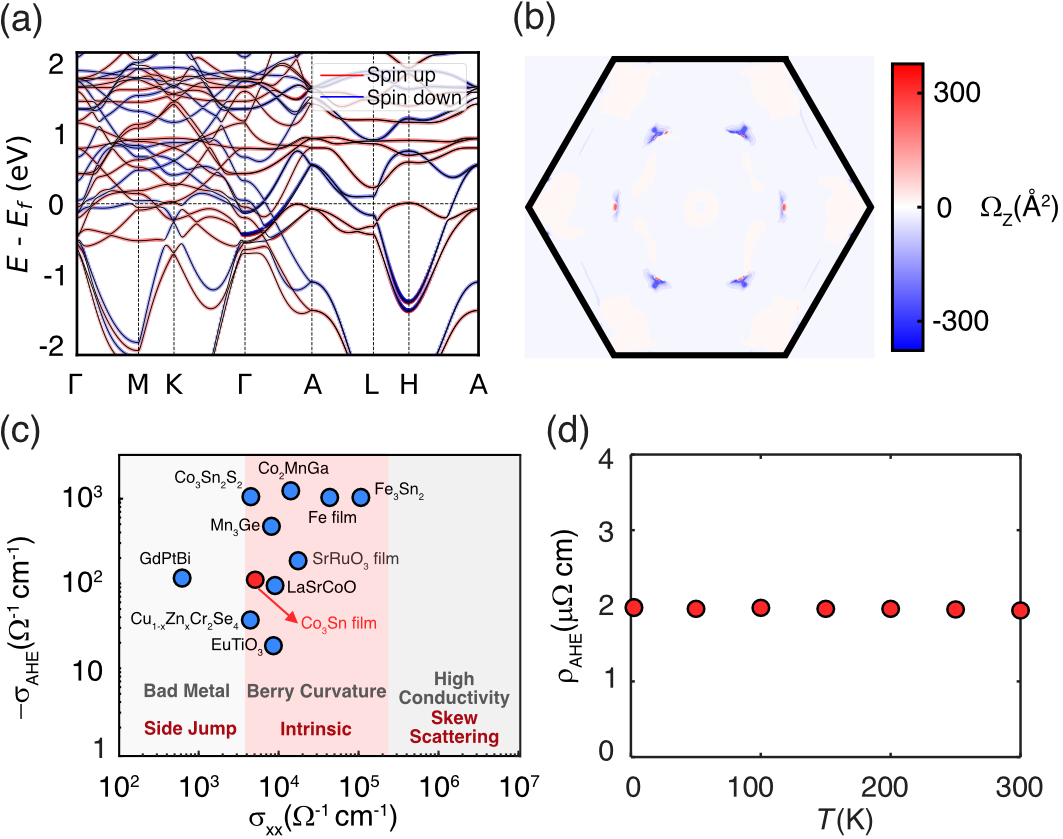}
\caption{\justifying\textbf{Berry curvature, band structure, and anomalous Hall conductivity.} Calculated band structure with SOC along high-symmetry paths in the Brillouin zone, where red and blue dots denote the spin projection along the $z$ direction. (b) Calculated Berry curvature $\Omega_z(\mathbf{k})$ distribution in the $k_z=0$ plane, showing intense hotspots around the Fermi-level. (c) Scaling analysis of $-\sigma_{\mathrm{AHE}}$ as a function of $\sigma_{xx}$ for various material systems~\cite{liu2018giant,manna2018colossal,chen2024colossal,nayak2016large,yu2000magnetotransport,skorupskii2024designing,miyasato2007crossover,takahashi2018anomalous,guin2019anomalous,singh2026berry}. The Co$_3$Sn film (red) falls into the intrinsic Berry curvature-dominated regime. (d) Temperature ($T$) dependence of $\rho_{\mathrm{AHE}}$ from 2 to 300\,K.}
\label{fig4}
\end{figure}

Our comprehensive materials characterization indicates that Co$_3$Sn is a metal with easy-plane ferromagnetism characterized by a sizable AHE ($\sigma_{\rm AHE}\approx\rho_{xy}/\rho_{xx}^2\approx90\,\Omega^{-1}{\rm cm}^{-1}$). To elucidate the microscopic origin of the AHE, we first performed density functional theory (DFT) calculations of the electronic band structure of Co$_{3}$Sn [Fig.~\ref{fig4}(a)]. The band structure reveals multiple gapped-out band crossings between spin-split bands near the Fermi level; however, our analysis reveals that distinct topological band structure features, such as Weyl points, are absent. Figure~\ref{fig4}(b) displays the Berry curvature $\Omega_z(\mathbf{k})$ calculated from DFT at the Fermi energy, shown as a map in the $k_z=0$ plane. Berry curvature hotspots can be detected halfway along the $\Gamma-K$ direction, consistent with the presence of a spin-polarized band below Fermi energy at this Brillouin zone position. The calculation of the AHE conductivity from the band structure using the Kubo formula yields $\sigma_\mathrm{AHE}^{\mathrm{calc}} = 108\,\Omega^{-1}\mathrm{cm}^{-1}$. 

This calculated value is in excellemt agreement with our experimentally determined value ($\sigma_{\rm AHE}\approx\rho_{xy}/\rho_{xx}^2\approx90\,\Omega^{-1}{\rm cm}^{-1}$), suggesting the AHE of Co$_3$Sn is originating from intrinsic Berry curvature. To further validate this understanding, we compare the AHE conductivity of Co$_{3}$Sn with values reported for various magnetic materials~\cite{liu2018giant,manna2018colossal,chen2024colossal,nayak2016large,yu2000magnetotransport,skorupskii2024designing,miyasato2007crossover,takahashi2018anomalous,guin2019anomalous,singh2026berry} in Fig.~\ref{fig4}(c). In the conventional \(-\sigma_{\mathrm{AHE}}\) versus \(\sigma_{xx}\) classification~\cite{nagaosa2010ahe}, we find that Co$_3$Sn falls into the intrinsic regime, where Berry curvature contribution should dominate the AHE signal owing to its moderate electric conductivity $\sigma_{xx}=\rho_{xx}^{-1}\approx6700\,\Omega^{-1}{\rm cm}^{-1}$ at room temperature. The intrinsic Berry curvature origin of the AHE of Co$_3$Sn is also consistent with the detected temperature independence of both $\rho_{\rm AHE}$ [Fig.~\ref{fig4}(d)] and $\rho_{xx}$. Owing to $\rho_{xy}\approx\sigma_{\rm AHE}\cdot\rho_{xx}^2$ for $\rho_{xx}\gg\rho_{xy}$ it follows that  $\sigma_{\rm AHE}$ is essentially temperature independent, as expected for Berry curvature contributions. We note in passing that the $T$-independence of $\sigma_{\rm AHE}$ indicates a Curie temperature of in-plane ferromagnetism much above 300\,K.

We close the discussion with a comparison of the electronic and magnetic properties of Co$_3$Sn and Co$_3$Sn$_2$S$_2$. Our results suggests that the different layer stacking sequence of these two compounds [see Fig.~\ref{fig1}(a)] results in markedly different electronic and magnetic properties. Both compounds are ferromagnets, but Co$_3$Sn$_2$S$_2$ develops an out-of-plane magnetization below 180\,K~\cite{liu2018giant} whereas Co$_3$Sn exhibits in-plane ferromagnetism above room temperature. While further measurements, also at higher temperature, and model calculation are required to determine the magnetic structure of Co$_3$Sn in more detail, its higher Curie temperature is intuitively consistent with the direct A-B stacking, promoting larger exchange coupling. Regarding the electronic band structure, Weyl points close to Fermi energy promote a giant AHE exceeding $10^3\,\Omega^{-1}{\rm cm}^{-1}$ in Co$_3$Sn$_2$S$_2$. On the other hand, our analysis shows that Co$_3$Sn lacks such distinct topological band structure features, resulting in a moderate AHE $(\sigma_{\rm AHE}\approx90\,\Omega^{-1}{\rm cm}^{-1})$.

\section{CONCLUSION}
\label{sec:conclusion}

In summary, we have successfully grown high-quality thin films of Co$_3$Sn using molecular beam epitaxy, a material that has not yet been synthesized in bulk crystal form. Our comprehensive characterization of the structural, magnetic, and electric thin film properties reveal metallic behavior and easy-plane ferromagnetism above room temperature. Our study reveals a temperature-independent anomalous Hall effect up to room temperature whose characteristics are in agreement with an intrinsic Berry curvature effect predicted by theoretical model calculations. Moreover, our results also highlight the sensitivity of magnetic and electronic properties of kagome metals to the layer stacking sequence, where direct A-B stacking of the kagome layers in Co$_3$Sn and Fe$_3$Sn~\cite{prodan2023large} seems to favor in-plane ferromagnetism with large Curie temperature. At the same time, our studies reaffirms the understanding that ferromagnetic kagome metals based on transition metals are a fertile ground for investigating Berry-curvature dominated magneto-transport effects~\cite{liu2018giant, wang2018large, ye2018massive, sankar2025room}. Looking ahead, our study opens pathways for tuning the magnetic and electronic properties of kagome metals within this material family~\cite{liu2018giant, wang2018large, kang2020topological, chen2023visualizing, chen2026nematic,ye2018massive, kang2020dirac, sankar2024experimental, sankar2025room} using atomic layer epitaxy and the synthesis of epitaxial heterostructures.

\begin{acknowledgments}
We thank Jiayin Tang and Danfeng Li for help with the device fabrication.
\end{acknowledgments}

\paragraph{Author Contributions}
 B.J. and S.S. conceived and designed the project. Y.Q. and S.S. performed the MBE growth and magneto-transport measurements and analyzed the data. Y.Q. and S.S. fabricated the Hall bar devices. Y.J. performed the magnetization measurements and X.C. conducted the DFT calculations and symmetry analysis. B.J. and S.S. wrote the manuscript with input from all authors.

\paragraph{Data Availability} Replication data for this manuscript are available via Zenodo https://zenodo.org/records/XXXXX.
\bibliography{Anisotropy}
\bibstyle{apsrev4-1}

\end{document}